\documentclass[iop,apj,a4paper]{emulateapj}
\usepackage{apjfonts,amsmath,amssymb,epsf,lscape,color,ulem}

\shortauthors{Taniguchi~et~al.}
\shorttitle{Massive,~mostly~star-formation~quenched~galaxies}

\begin{document}


\title{Discovery of Massive, Mostly Star-formation Quenched Galaxies
with Extremely Large Lyman-alpha Equivalent Widths at $z \sim
3$\altaffilmark{*}}

\author{
Yoshiaki~Taniguchi\altaffilmark{1},
Masaru~Kajisawa\altaffilmark{1,2},
Masakazu~A.~R.~Kobayashi\altaffilmark{1},
Tohru~Nagao\altaffilmark{1},
Yasuhiro~Shioya\altaffilmark{1},
Nick~Z.~Scoville\altaffilmark{3},
David~B.~Sanders\altaffilmark{4},
Peter~L.~Capak\altaffilmark{3,5},
Anton~M.~Koekemoer\altaffilmark{6},
Sune~Toft\altaffilmark{7},
Henry~J.~McCracken\altaffilmark{8},
Olivier~Le~F\`evre\altaffilmark{9},
Lidia~Tasca\altaffilmark{9},
Kartik~Sheth\altaffilmark{10},
Alvio~Renzini\altaffilmark{11},
Simon~Lilly\altaffilmark{12},
Marcella~Carollo\altaffilmark{12},
Katarina~Kova\v{c}\altaffilmark{12},
Olivier~Ilbert\altaffilmark{9},
Eva~Schinnerer\altaffilmark{13},
Hai~Fu\altaffilmark{14},
Laurence~Tresse\altaffilmark{9},
Richard~E.~Griffiths\altaffilmark{15}, and
Francesca~Civano\altaffilmark{16}
}

\email{tani@cosmos.phys.sci.ehime-u.ac.jp}

\altaffiltext{*}{Based on observations with NASA/ESA \textit{Hubble
Space Telescope}, obtained at the Space Telescope Science Institute,
which is operated by AURA, Inc., under NASA contract NAS 5-26555; also
based on data collected at the Subaru Telescope, which is operated by
the National Astronomical Observatory of Japan; and also based on data
products from observations made with ESO Telescopes at the La Silla
Paranal Observatory under ESO programme ID 179.A-2005 and on data
products produced by TERAPIX and the Cambridge Astronomy Survey Unit
on behalf of the UltraVISTA consortium.}

\altaffiltext{1}{Research Center for Space and Cosmic Evolution, Ehime
University, Bunkyo-cho, Matsuyama 790-8577, Japan}

\altaffiltext{2}{Graduate School of Science and Engineering, Ehime
University, Bunkyo-cho, Matsuyama 790-8577, Japan}

\altaffiltext{3}{Department of Astronomy, MS 105-24, California
Institute of Technology, Pasadena, CA 91125, USA}

\altaffiltext{4}{Institute for Astronomy, University of Hawaii, 2680
Woodlawn Drive, Honolulu, HI 96822, USA}

\altaffiltext{5}{Spitzer Science Center, California Institute of
Technology, Pasadena, CA 91125, USA}

\altaffiltext{6}{Space Telescope Science Institute, 3700 San Martin
Drive, Baltimore, MD 21218, USA}

\altaffiltext{7}{Dark Cosmology Centre, Niels Bohr Institute,
University of Copenhagen, Juliane Mariesvej 30, DK-2100 Copenhagen,
Denmark}

\altaffiltext{8}{Institut d'Astrophysique de Paris, UMR7095 CNRS,
Universit\'{e} Pierre et Marie Curie, 98 bis Boulevard Arago, 75014,
Paris, France}

\altaffiltext{9}{Aix Marseille Universit\'{e}, CNRS, LAM (Laboratoire
d'Astrophysique de Marseille), UMR 7326, 13388, Marseille, France}

\altaffiltext{10}{National Radio Astronomy Observatory, 520 Edgemont
Road, Charlottesville, VA 22903, USA}

\altaffiltext{11}{Dipartimento di Astronomia, Universita di Padova,
vicolo dell'Osservatorio 2, 35122, Padua, Italy}

\altaffiltext{12}{Department of Physics, ETH Zurich, 8093, Zurich,
Switzerland}

\altaffiltext{13}{MPI for Astronomy, K\"{o}nigstuhl 17, D-69117
Heidelberg, Germany}

\altaffiltext{14}{Department of Physics \& Astronomy, University of
Iowa, Iowa City, IA 52245, USA}

\altaffiltext{15}{Physics \& Astronomy, STB-216, U. Hawaii at Hilo,
Hilo, HI 96720, USA}

\altaffiltext{16}{Yale Center for Astronomy and Astrophysics, 260
Whitney Avenue, New Haven, CT 06520, USA; Smithsonian Astrophysical
Observatory, 60 Garden Street, Cambridge, MA 02138, USA}

\begin{abstract}

 We report a discovery of 6 massive galaxies with both extremely large
 Ly$\alpha$ equivalent width and evolved stellar population at
 $z\sim3$.  These MAssive Extremely STrong Ly$\alpha$ emitting Objects
 (MAESTLOs) have been discovered in our large-volume systematic survey
 for strong Ly$\alpha$ emitters (LAEs) with twelve optical
 intermediate-band data taken with Subaru/Suprime-Cam in the COSMOS
 field.  Based on the SED fitting analysis for these LAEs, it is found
 that these MAESTLOs have (1)~large rest-frame equivalent width of
 $EW_0 (\mathrm{Ly}\alpha)\sim100$--300~{\AA},
 (2)~$M_\star\sim10^{10.5}$--$10^{11.1}~M_\odot$, and (3)~relatively
 low specific star formation rates of
 $SFR/M_\star\sim0.03$--$1~\mathrm{Gyr}^{-1}$.  Three of the 6
 MAESTLOs have extended Ly$\alpha$ emission with a radius of several
 kpc although they show very compact morphology in the HST/ACS images,
 which correspond to the rest-frame UV continuum.  Since the MAESTLOs
 do not show any evidence for AGNs, the observed extended Ly$\alpha$
 emission is likely to be caused by star formation process including
 the superwind activity.  We suggest that this new class of LAEs,
 MAESTLOs, provides a missing link from star-forming to passively
 evolving galaxies at the peak era of the cosmic star-formation
 history.

\end{abstract}


 \section{INTRODUCTION}

 Most of massive galaxies in the present universe are passively
 evolving galaxies with little on-going star formation (e.g.,
 \citealp{kau03}).  In the current understanding of galaxy evolution,
 massive galaxies are considered to have evolved more rapidly than
 less massive systems in earlier universe: so-called the downsizing
 evolution of galaxies \citep{cow96}.  These massive galaxies have
 formed their stars actively by a cosmic age of a few Gyr (redshift
 $z\sim2$--3), when the cosmic star formation rate density peaked
 (e.g., \citealp{bou11}).  After this epoch, their star formation
 stopped and they passively evolved into elliptical galaxies seen
 today.  However, the quenching mechanism of star formation in these
 massive galaxies has not yet been understood because the process may
 have occurred in a relatively short time scale, making it difficult
 to observe such events (e.g., \citealp{ren09}; \citealp{pen10};
 \citealp{dur15}; \citealp{man15}).

 To seek for star-forming galaxies in young universe, the Hydrogen
 Ly$\alpha$ emission provides the most useful tool.  Therefore, many
 searches for redshifted Ly$\alpha$ emission have resulted in the
 discovery of young galaxies beyond $z\sim7$, corresponding to an
 cosmic age of $\lesssim 750$~Myr (\citealp{ono12}; \citealp{shi12};
 \citealp{fin13}; \citealp{kon14}; \citealp{sch14}).  Among such
 Ly$\alpha$ emitting galaxies (Ly$\alpha$ emitters, hereafter LAEs),
 those with a very large equivalent width (EW), i.e., extremely strong
 LAEs, are particularly important in that they can be galaxies in a
 very early stage of galaxy formation (e.g., \citealp{sch03};
 \citealp{nag07}).

 \begin{deluxetable*}{clccccccc}
  \tablecaption{Physical Properties of the 6 MAESTLOs\label{tab1}}
  \tablewidth{\linewidth}
  \tablehead{
  \colhead{No.} &
  \colhead{$z_\mathrm{phot}$} &
  \colhead{$\log{M_\star}$} &
  \colhead{$\tau$}       &
  \colhead{age}          &
  \colhead{$E(B-V)$}     &
  \colhead{$EW_0(\mathrm{Ly}\alpha)$} &
  \multicolumn{2}{c}{$\log{[SFR/(M_\odot~\mathrm{yr^{-1}})]}$} \\
  \cline{8-9} \vspace{-2mm}\\
  \colhead{} &
  \colhead{} &
  \colhead{$(M_\odot)$} &
  \colhead{(Gyr)}  &
  \colhead{(Gyr)}  &
  \colhead{(mag)} &
  \colhead{(\AA)} &
  \colhead{Ly$\alpha$} &
  \colhead{SED}
  }
  \startdata
 1 & 3.16                  & $11.11^{+0.08}_{-0.00}$ & $0.32^{+0.08}_{-0.00}$ & $1.61^{+0.29}_{-0.18}$ & $0.03^{+0.06}_{-0.00}$ & $240^{+20}_{-19}$ & $1.12\pm0.02$ & $0.63^{+0.26}_{-0.00}$ \vspace{1mm}\\
 2 & 2.81                  & $11.11^{+0.04}_{-0.07}$ & $1.59^{+1.58}_{-0.59}$ & $1.80^{+0.40}_{-0.52}$ & $0.29^{+0.03}_{-0.02}$ & $306\pm20$      & $1.20\pm0.02$ & $1.79^{+0.12}_{-0.08}$ \vspace{1mm}\\
 3 & 2.81\tablenotemark{a} & $10.90^{+0.00}_{-0.00}$ & $0.05^{+0.00}_{-0.00}$ & $0.29^{+0.00}_{-0.00}$ & $0.19^{+0.00}_{-0.00}$ & $172\pm6$       & $1.04\pm0.02$ & $0.88^{+0.00}_{-0.00}$ \vspace{1mm}\\
 4 & 3.24                  & $10.71^{+0.12}_{-0.03}$ & $0.40^{+0.40}_{-0.00}$ & $1.28^{+0.62}_{-0.14}$ & $0.12^{+0.05}_{-0.04}$ & $178\pm16$      & $0.88\pm0.03$ & $0.94^{+0.22}_{-0.16}$ \vspace{1mm}\\
 5 & 2.50\tablenotemark{b} & $10.54^{+0.08}_{-0.04}$ & $0.06^{+9.94}_{-0.02}$ & $0.14^{+0.26}_{-0.03}$ & $0.40^{+0.04}_{-0.03}$ & $107^{+11}_{-14}$ & $0.68\pm0.03$ & $1.94^{+0.37}_{-0.27}$ \vspace{1mm}\\
 6 & 3.16                  & $10.52^{+0.03}_{-0.03}$ & $0.50^{+0.00}_{-0.00}$ & $1.90^{+0.00}_{-0.10}$ & $0.02^{+0.02}_{-0.01}$ & $124\pm16$      & $0.64\pm0.05$ & $0.41^{+0.08}_{-0.03}$
  \enddata

  \tablecomments{The No. is given in order of decreasing the estimated
  stellar mass.  Errors for the quantities correspond to 1-$\sigma$
  confidence interval (i.e., $\Delta\chi^2\le1$) estimated from the
  SED fitting.  In the SED fitting, the templates older than cosmic
  age at $z_\mathrm{phot}$ are not used.  The entry of 0.00 for these
  errors indicate that there is no parameter grid in
  $\Delta\chi^2\le1$ around the best-fit model parameter.}

  \tablenotetext{a}{$z_\mathrm{spec}=2.798$.}

  \tablenotetext{b}{$z_\mathrm{spec}=2.513$.}

 \end{deluxetable*}

 In order to search for them, we have carried out a survey for
 extremely strong LAEs over an unprecedentedly large volume.  While
 most of the detected objects with strong Ly$\alpha$ seem to be young
 galaxies with small stellar mass as expected for LAEs, we have
 serendipitously found 6 massive galaxies with extremely large $EW
 (\mathrm{Ly}\alpha)$ and relatively evolved stellar population at
 $z\sim3$, that show no evidence for an active galactic nucleus (AGN).
 Here we present the physical properties of this new population that
 is expected to be in a transition phase between star-forming and
 passive evolution.  In this Letter, we use a standard cosmology with
 $\Omega_M=0.3$, $\Omega_\Lambda=0.7$, and
 $H_0=70~\mathrm{km~s^{-1}~Mpc^{-1}}$.

 \section{Data and Analysis}

 \begin{figure*}
  \epsscale{1.094}
  \plotone{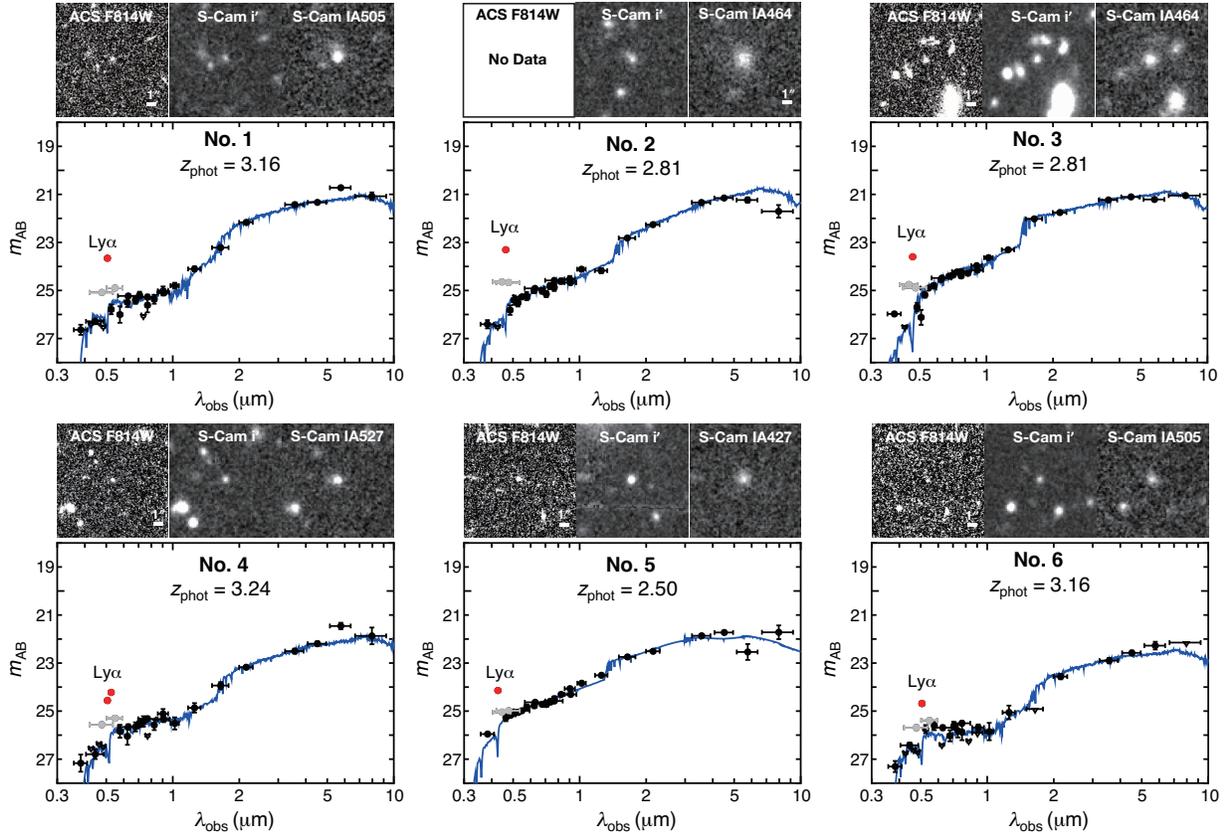}

  \caption{SEDs and HST and Subaru images of the 6 MAESTLOs.  In the
  bottom main panel, the observed data points are shown by filled
  circles with error bars.  The red symbol shows the IA band with a
  significant flux excess by the Ly$\alpha$ emission, while the grey
  symbols represent the broad bands whose wavelength coverage overlaps
  with the excess IA band.  The blue curve shows the best-fit model
  SED from the GALAXEV library.  The data affected by the strong
  Ly$\alpha$ emission (red and grey points) are not used in the SED
  fitting.  The inverted triangles represents the $3\sigma$ upper
  limit for the undetected bands.  In the top panels, the thumbnails
  of the HST ACS in the F814W filter, the Subaru Suprime-Cam
  $i^\prime$-band image, and the excess IA-band image are shown.  Each
  panel is $12^{\prime\prime}\times12^{\prime\prime}$ in size.  The
  ACS images are convolved with a Gaussian kernel with $\sigma=1$~pix
  ($=0\farcs03$) to reduce the noise.\label{fig1}}

 \end{figure*}

 In this study, we use the multi-wavelength data set from the Cosmic
 Evolution Survey (COSMOS; \citealp{sco07}).  Optical imaging data
 with 12 intermediate-band (hereafter, IA-band) filters equipped on
 Subaru/Suprime-Cam allow us to pick up strong emission-line objects
 by a significant flux excess in one of the IA bands.  The spectral
 resolution of our IA filters is $R=\lambda/\Delta\lambda=20$--26, and
 the 12 IA filters cover the whole optical wavelength range from
 4270~{\AA} to 8270~{\AA} (\citealp{tan15}).  Therefore, we can search
 for strong LAEs at $2.5<z<5.8$.  Although the details of our
 selection procedure for strong LAEs are given elsewhere
 (M.~A.~R.~Kobayashi et al., in prep.), we briefly summarize it in the
 following.

 At first, from the COSMOS Official Photometric Catalog (version 2012,
 \citealp{cap07}), we selected objects with a significant ($3\sigma$)
 flux excess in an IA band from the frequency-matched continuum
 estimated by using two adjacent broad-band magnitudes.  In order to
 identify which emission line causes the IA-band excess of these
 objects, we applied the public photometric redshift code EAZY
 \citep{bra08} to the multi-band photometric data from optical to MIR,
 which include CFHT $u^*$ and $i^*$, Subaru $Bg'Vr'i'z'$ and 12 IA
 bands \citep{tan07}, UltraVISTA $YJHK$ \citep{mcc12}, and
 Spitzer/IRAC 3.6~{\micron} and 4.5~{\micron} bands \citep{san07}.
 The excess IA band and any broad bands whose wavelength coverage is
 overlapped with the excess IA band are excluded from the photometric
 redshift calculation.  We adopted a line identification with the
 highest probability in the volume-weighted redshift likelihood
 function, and assigned the photometric redshift assuming the emission
 line enters into the effective wavelength of the excess IA band.  We
 selected LAEs from these strong emission-line objects and then
 performed the spectral energy distribution (SED) fitting with the
 GALAXEV population synthesis model (\citealp{bru03}) to estimate the
 physical properties of the LAEs.  In the SED fitting, we assumed the
 exponentially decaying star formation histories with an $e$-folding
 timescale of $\tau=0.01$--10~Gyr.  The Chabrier initial mass function
 \citep{cha03} and the Calzetti extinction law \citep{cal00} were
 adopted.  The excess IA band and any broad bands overlapping with the
 excess band were again excluded. Although other strong emission lines
 such as [\ion{O}{2}], [\ion{O}{3}], and H$\alpha$ may enter into the
 $JHK$ bands, we used all $JHK$-band data in the fitting, because the
 effect of such emission lines is not expected to be serious for these
 bands with wide filter bandpasses.  In addition to the multi-band
 photometry used in the photometric redshift estimate, we also used
 the IRAC 5.8~{\micron} and 8.0~{\micron} bands to obtain more
 accurate physical properties such as the stellar mass and age.  Our
 survey covers a 1.34~$\mathrm{deg}^2$ area in the COSMOS field, that
 is, the overlapped area between COSMOS deep region and UltraVISTA DR1
 \citep{mcc12}.  The wide survey area and wide wavelength coverage of
 the 12 IA bands allow us to search for strong LAEs at $2.5<z<5.8$
 over a very large volume of $5.5\times10^7~\mathrm{Mpc}^3$.

 As a result, we obtain a sample of 589 LAEs at $2.5<z<5.3$.  In this
 sample, 18 LAEs have both an extremely large rest-frame equivalent
 width of $EW_0(\mathrm{Ly}\alpha)\ge100$~{\AA} and a large stellar
 mass with $M_\star\ge10^{10.5}~M_\odot$.  Hereafter, we call these 18
 LAEs MAssive, Extremely STrong Ly$\alpha$ emitting Objects
 (MAESTLOs).  Since our main interest is the star-forming activity in
 galaxies, we rejected possible active galactic nuclei (AGN) by using
 the IRAC color criteria proposed by \citet{don12}.  We also used the
 XMM-COSMOS \citep{has07}, Chandra-COSMOS (\citealp{elv09};
 \citealp{civ12}) and Chandra COSMOS Legacy (F.~Civano et al., in
 prep.) and VLA-COSMOS (\citealp{sch07}) catalogues to reject AGNs.
 In total, 12 MAESTLOs turn out to show evidence for AGN.
 Accordingly, we obtain a sample of 6 MAESTLOs without evidence for
 AGN\footnote{Note that a stacking analysis for the 6 MAESTLOs,
 corresponding to a $\sim650$~ksec exposure, results in no detection.
 The 95\% upper limit in the 0.5--2~keV band is
 $3.63\times10^{-5}~\mathrm{counts~s^{-1}}$ which corresponds to a
 rest-frame luminosity of $7.76\times10^{42}~\mathrm{ergs~s^{-1}}$ at
 $z\sim3$.}.  Their observational properties are summarized in
 Table~\ref{tab1}.

 Their sizes are measured in the excess IA-band (i.e., Ly$\alpha$
 image) and the COSMOS HST/ACS $I_{\rm F814W}$-band mosaics
 \citep{koe07}, corresponding to the rest-frame UV continuum, by using
 the GALFIT code \citep{pen02}.  We fit the observed surface
 brightness with an exponential law, taking account of the point
 spread function (PSF) of these data.  Here we fix the S{\'e}rsic
 index to $n=1$ because our data are not deep enough to resolve the
 degeneracy between the radius and S{\'e}rsic index for our MAESTLOs.
 The PSF images of the excess IA-band and ACS data are measured by
 combining relatively bright isolated stars in each image.  Note,
 however, that we cannot measure the ACS sizes for MAESTLO No.~2 since
 this object is out of the HST/ACS field.  Furthermore, we also
 measure the sizes of the rest-frame UV continuum using the
 $i^\prime$-band data taken with the same instrument as that of the
 excess IA-band data.  Note that the half light radius of the PSF in
 the excess IA bands is 0.75--0.83~arcsec, while that in the
 $i^\prime$ band is 0.51~arcsec.  In order to estimate the uncertainty
 in the size measurements including the systematic effects such as the
 background fluctuation, we carry out the Monte Carlo simulation as in
 previous studies (e.g., \citealp{str15}).  After adding the best-fit
 model profile to the image at 200 random positions around the
 original position (in a $2^\prime\times2^\prime$ region), we
 re-measure their sizes.  The standard deviation of these 200
 measurements is adopted as the size uncertainty.  In order to check
 whether the object is significantly extended or not, we also
 calculated the fraction of the cases that GALFIT returned the
 ``unresolved'' flag, $f_\mathrm{unres}$, in the 200
 measurements.  The estimated half-light radii and errors of the
 MAESTLOs together with $f_\mathrm{unres}$ are given in
 Table~\ref{tab2}.

 We have also carried out additional simulations.  In these
 simulations, (1)~we convolved model galaxies with the best-fit
 F814W-band light profile with the PSF of the IA-band data, (2)~then,
 we added them into the IA-band image, and (3)~we measured their sizes
 with GALFIT.  We performed 200 such simulations for the two MAESTLOs
 with the extended Lya emission (Nos.~1 and 3) for which ACS image is
 available; note that No.~2 is located out of the ACS coverage.  We
 then find that GALFIT returned the ``unresolved'' flag in most cases
 (185/200 and 169/200 for Nos.~1 and 3, respectively).  Therefore, we
 have confirmed that their Ly$\alpha$ emission is really extended in
 their IA images.  In conclusion, the three MAESTLOs have an extended
 Ly$\alpha$ emission with a size of several~kpc.

 \begin{deluxetable}{ccccccc}
  \tablecaption{Size measurements of the 6 MAESTLOs \label{tab2}}
  \tablewidth{\linewidth}
  \tablehead{
  \colhead{} &
  \multicolumn{2}{c}{ACS F814W} &
  \multicolumn{2}{c}{S-Cam $i^\prime$} &
  \multicolumn{2}{c}{S-Cam IA}\\
  \cline{2-3} \cline{4-5} \cline{6-7}\vspace{-2mm}\\
  \colhead{No.} & $r_\mathrm{HL}$~(kpc) & $f_\mathrm{unres}$ & $r_\mathrm{HL}$~(kpc) & $f_\mathrm{unres}$ & $r_\mathrm{HL}$~(kpc) & $f_\mathrm{unres}$
  }
  \startdata
 1 & $0.52\pm0.08$ & 0.060 & $<3.87$\tablenotemark{b} & 0.660 & $4.50\pm0.61$\vspace{1mm}& 0 \\
 2 & ---\tablenotemark{a} & ---\tablenotemark{a} & $4.49\pm2.01$ & 0.175 & $6.68\pm0.70$\vspace{1mm}& 0 \\
 3 & $1.00\pm0.09$ & 0 & $4.21\pm1.75$ & 0.115 & $7.18\pm0.80$\vspace{1mm}& 0 \\
 4 & $0.34\pm0.12$ & 0.100 & $<3.83$\tablenotemark{b} & 0.750 & $<6.20$\tablenotemark{b}\vspace{1mm} & 0.685 \\
 5 & $<0.57$\tablenotemark{b} & 0.880 & $<4.12$\tablenotemark{b} & 0.710 & $ 3.77\pm1.83$\vspace{1mm}& 0.050 \\
 6 & $<0.54$\tablenotemark{b} & 0.645 & $<3.87$\tablenotemark{b} & 0.670 & $<5.74$\tablenotemark{b}& 0.410
  \enddata

  \tablecomments{$r_\mathrm{HL}$ is half-light radius and
  $f_\mathrm{unres}$ is Errors for $r_\mathrm{HL}$ are based on the
  Monte Carlo simulation described in the text.}

  \tablenotetext{a}{Out of the ACS/F814W-band data.}

  \tablenotetext{b}{Unresolved.}

 \end{deluxetable}
 \section{Results \& Discussions}

 In Figure~\ref{fig1}, we show the rest-frame UV--NIR SED of the 6
 MAESTLOs together with their thumbnails in the excess IA, $i^\prime$,
 and ACS $I_{\rm F814W}$ bands.  It is found that they are
 significantly bright in the rest-frame NIR wavelengths, leading to
 their large estimated stellar masses of
 $\log{(M_\star/M_\odot)}=10.5$--11.1.  Another unexpected property is
 that they show very red rest-frame UV-optical colors despite of their
 extremely large $EW_0(\mathrm{Ly}\alpha)$; i.e., the MAESTLOs show a
 relatively strong 4000~{\AA} continuum break in the rest-frame
 optical as well as the Lyman break in the rest-frame far-UV.  These
 continuum features allow us to identify the flux excess in a
 concerned IA band as the Ly$\alpha$ emission line, resulting in an
 accurate photometric redshift for them.  In fact, two of the 6
 MAESTLOs have spectroscopic identifications and their spectroscopic
 redshifts agree with the photometric redshifts estimated from the
 IA-band excess (Nos.~3 and 5; see Table~1).  The strong 4000~{\AA}
 break observed in the MAESTLOs suggests relatively old stellar
 population in them, and their best-fit stellar ages based on SED
 fitting are 1--2~Gyr\footnote{Note that 2 MAESTLOs (i.e., Nos.~3 and
 5) have relatively young ages.  No.~3 has a very short $e$-folding
 timescale and a clear Balmer break in the SED, suggesting that its
 SFR is rapidly decreasing.  On the other hand, No.~5 shows relatively
 weak Balmer/4000~{\AA} break compared to the other MAESTLOs and its
 $e$-folding timescale is highly uncertain.}.  Thus these galaxies
 form a completely different population from typical high-redshift
 LAEs with small stellar masses and young stellar ages (e.g.,
 \citealp{ono10}; \citealp{hag14}).

\begin{figure}
  \plotone{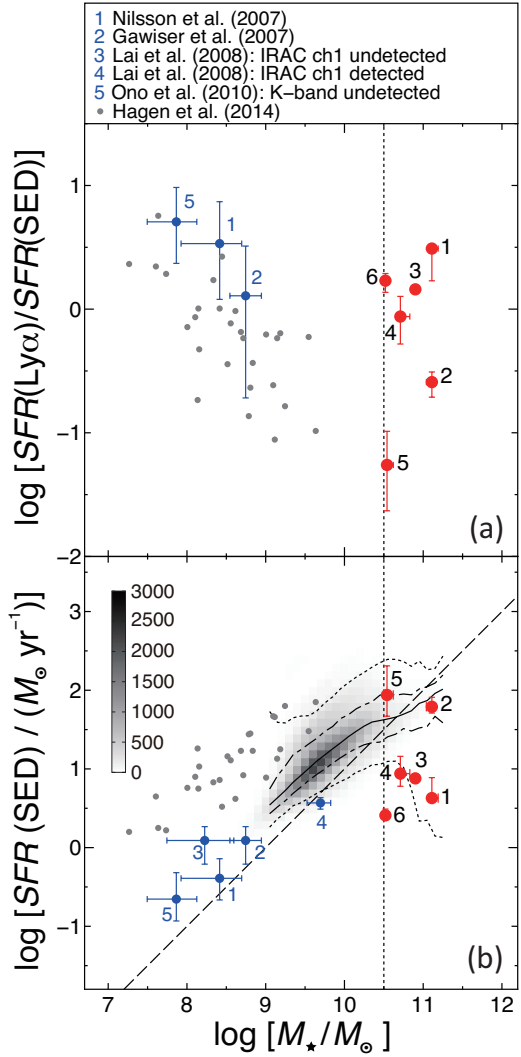}

  \caption{(a) $SFR(\mathrm{Ly}\alpha)/SFR(\mathrm{SED})$
  vs. $M_\star$ and (b) $SFR(\mathrm{SED})$ vs. $M_\star$ for our
  MAESTLOs (red circles) and LAEs at $z\sim3$ from the literatures
  (blue and gray circles for the results from stacked LAEs and from
  individual LAE, respectively).  For \citet{hag14}, we plot only the
  LAEs at $z=2.5$--3.2 in their sample.  Note that the SFRs and
  $M_\star$ in the literatures derived with the Salpeter IMF are
  converted into those with the Chabrier IMF we use.  Note also that
  \citet{hag14} used the SFRs estimated from the dust-corrected UV
  luminosity instead of those based on the SED fitting.  In panel (b),
  we also plot the so-called main sequence of star-forming galaxies at
  $z_\mathrm{phot}=2.5$--3.2 in the COSMOS field in gray scale with
  bin size of 0.1~dex in $M_\star$ and SFR.  The scale (the number of
  galaxies per bin) is shown in the upper-left inset.  Solid,
  dot-dashed, and dotted lines show the median, 16 and 84 percentiles,
  and 2.5 and 97.5 percentiles in bins of 0.1~dex in stellar mass.
  The dashed line shows the relation of $sSFR=1~\mathrm{Gyr}^{-1}$.
  The vertical dotted line corresponds to
  $\log{(M_\star/M_\odot)}=10.5$, which is one of the criteria of
  MAESTLO.\label{fig2}}
 \end{figure}

 Despite their relatively old stellar population, the MAESTLOs have
 extremely large $EW_0(\mathrm{Ly\alpha})$ of $\sim 100$--300~{\AA}.
 In order to compare the star formation rates (SFR) estimated from
 Ly$\alpha$ luminosity, $SFR(\mathrm{Ly\alpha})$, with that from SED
 fitting, $SFR(\mathrm{SED})$, we show
 $SFR(\mathrm{Ly\alpha})/SFR(\mathrm{SED})$ ratios of the MAESTLOs as
 a function of stellar mass in Figure~\ref{fig2}~(a).  Here, we use
 the \citet{ken98} relation between $SFR$ and $L(\mathrm{H\alpha})$
 combined with both the $L(\mathrm{Ly\alpha})/L(\mathrm{H\alpha})$
 ratio of 8.7 under the case B recombination and a correction factor
 converting from the Salpeter IMF into the Chabrier IMF (i.e.,
 multiplied by a factor of 0.60).

 For typical LAEs, it is found that the $SFR$ ratio decreases with
 increasing stellar mass \citep{hag14}.  On the other hand, for most
 MAESTLOs, $SFR (\mathrm{Ly\alpha})$ is comparable to
 $SFR(\mathrm{SED})$ and thus their $SFR$ ratios are similar to those
 of typical LAEs with much smaller masses.  Therefore, it is suggested
 that the escape fraction of the Ly$\alpha$ emission is relatively
 high in these galaxies and/or that there are other additional energy
 sources besides the photoionization by massive OB stars.

 In order to investigate their evolutionary stage, we show the
 distribution of MAESTLOs in the $SFR (\mathrm{SED})$--$M_\star$ plane
 together with typical LAEs at $z\sim3$ (\citealp{nil07};
 \citealp{gaw07}; \citealp{lai08}; \citealp{ono10}; \citealp{hag14})
 and galaxies at $z_\mathrm{phot}=2.5$--3.2 in the COSMOS field
 (Figure~\ref{fig2}~(b)).  Compared to normal
 star-forming galaxies on the main sequence at similar stellar masses
 and redshifts, the MAESTLOs have a smaller specific SFR\footnote{One
 exception is MAESTLO No.~5, which has a weak Balmer break as
 mentioned above.  Its sSFR is consistent with the main sequence at
 the redshift.}, $sSFR=SFR/M_\star\sim0.03$--$1~\mathrm{Gyr}^{-1}$,
 suggesting that their star formation activities are just ceasing and
 that they are in a transition phase from actively star-forming into
 quiescent galaxies.  This contrasts with normal LAEs that tend to
 have a sSFR similar with or higher than main-sequence galaxies (e.g.,
 \citealp{hag14}).

\begin{figure}
  \plotone{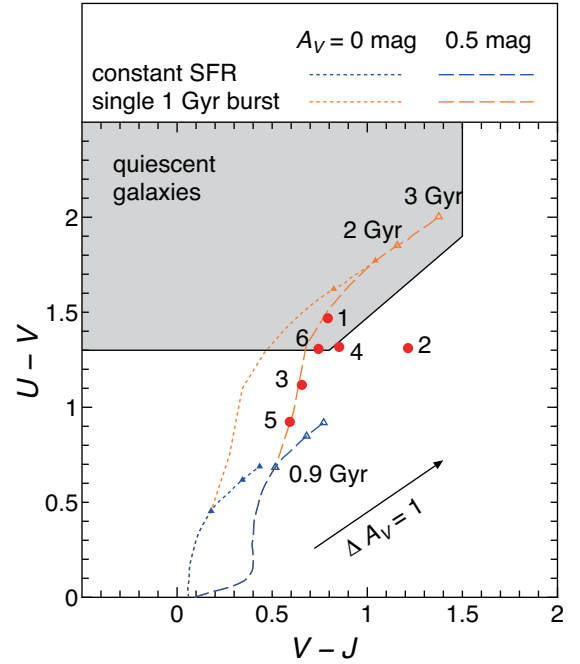}

  \caption{Rest-frame $U-V$ vs. $V-J$ two-color diagram for the
  MAESTLOs.  The 6 MAESTLOs are shown by filled circles with their ID
  numbers.  The domain of quiescent galaxies studied by \citet{muz13}
  is shown with the gray shaded area.  Two star formation models (the
  constant star formation and the single 1~Gyr starburst) are also
  shown by blue and orange dotted curves for $A_V=0$ and by blue and
  orange dashed curves for $A_V=0.5$ mag.\label{fig3}}

 \end{figure}

 In order to complementarily investigate the star formation histories
 of the MAESTLOs, we show the rest-frame $U-V$ vs. $V-J$ diagram in
 Figure~\ref{fig3}.  Comparing the quiescent galaxies studied by
 \citet{muz13}, we find that our four MAESTLOs with a low sSFR
 (Nos.~1, 3, 4, and 6) are located around the selection boundary for
 the quiescent galaxies and that their colors are consistent with the
 model tracks where star formation has been recently quenched.
 Therefore, this color analysis reinforces our scenario, suggesting
 that they have been recently quenched and are moving into the passive
 evolution phase.  Although the colors of the other two MAESTLOs are
 consistent with the star-forming models, their colors can also be
 interpreted as a galaxy that ceased its star formation recently.  The
 larger dust contents in these two galaxies may be expected if they
 are in an early phase of the superwind activity; i.e., most of dust
 grains may have not yet been blown out by the superwind.  We thus
 infer that MAESTLOs are in the final stage of massive galaxy
 formation where their SFRs decrease as gas is ejected from the galaxy
 as the superwind.

 As shown in Table~\ref{tab2}, the sizes in the rest-frame UV
 continuum of MAESTLOs are small (i.e., $\leq 1$~kpc).  It is
 noteworthy that their sizes are very similar to those of compact
 massive quiescent galaxies found at $z\sim2$ \citep{van14}, implying
 that the MAESTLOs can be interpreted as their progenitors.  It has
 been recently suggested that massive compact star-forming galaxies at
 $z\sim2$--3 evolve into compact quiescent galaxies after the cease of
 their star formation (e.g., \citealp{bar13}).  Although they are
 mostly dusty galaxies whose sizes are as small as the MAESTLOs, they
 have a younger age of $1.1^{+0.2}_{-0.6}$~Gyr and a higher sSFR of
 0.3--$3~\mathrm{Gyr}^{-1}$ than MAESTLOs \citep{bar14}.  We therefore
 suggest that they will evolve to passive galaxies through the MAESTLO
 phase.

 Our survey volume ($z=2.42$--2.59 and 2.72--3.42) corresponds to
 $1.4\times10^7~\mathrm{Mpc}^3$ and the number density of MAESTLOs is
 $4.3\times10^{-7}~\mathrm{Mpc}^{-3}$
 ($2.9\times10^{-7}~\mathrm{Mpc}^{-3}$ if we exclude MAESTLOs Nos.~2
 and 5 with a relatively high sSFR).  Thus MAESTLOs may have been
 missed by previous narrow-band surveys because their survey volumes
 were insufficient even in a survey with powerful instruments such as
 Subaru/Suprime-Cam (e.g., \citealp{ouc08}).

 Here we compare the number density of MAESTLOs with star-forming and
 quiescent galaxies with $M_\star>10^{10.5}~M_\odot$ using the stellar
 mass function of galaxies at the same redshift range \citep{ilb13}.
 We then find that the MAESTLOs constitute only 0.2--0.6\% of
 star-forming galaxies and $\sim2$\% of quiescent galaxies.  If we
 assume that all galaxies with $M_\star>10^{10.5}~M_\odot$ pass the
 phase of MAESTLO when they evolve from star-forming to quiescent
 galaxies, we obtain the duration of this phase, $\sim0.02\times
 t_\mathrm{univ}(z\sim\mbox{3--4})\sim30$--50~Myr ($\sim20$--34~Myr if
 we exclude Nos.~2 and 5), making them a rare population.  Such a
 short timescale truncation has been recently discussed based on other
 observational properties of galaxies at $z\sim3$ (\citealp{dur15};
 \citealp{man15}).

 Finally, we mention about the extended nature of Ly$\alpha$ emission.
 As demonstrated in Section 2, three of the 6 MAESTLOs (Nos.~1, 2, and
 3) have extended Ly$\alpha$ emission in the excess IA band, while all
 the MAESTLOs are compact in the both ACS $I_{\rm F814W}$-band and
 Subaru $i^\prime$-band data (see Table~\ref{tab2}).  The half-light
 radius of these three MAESTLOs in the IA-band data is $\sim4$--7~kpc,
 while that in the ACS data is $\leq1$~kpc for all the MAESTLOs
 (Table~\ref{tab2}).  This extended nature of Ly$\alpha$ emission is
 intimately related to the observed extremely large
 $EW_0(\mathrm{Ly}\alpha)$.  Plausible origins of the extended
 Ly$\alpha$ emission is attributed to (1)~scattering of Ly$\alpha$
 photon supplied from the central region of each MAESTLO,
 (2)~photoionized gas by massive star in the central region of each
 MAESTLO, or (3)~shock-heated gas driven by a superwind.  In addition,
 there are other possible ideas to explain the observed extended
 Ly$\alpha$ emission.  The first idea is a projection effect by a
 nearby LAE with a MAESTLO.  However, we consider that such a
 projection effect cannot be the origin of most MAESTLOs because of
 the small number densities of both massive galaxies with relatively
 low sSFR and LAEs with extremely large EWs at $z\sim3$ (e.g.,
 $\sim1.3\times10^{-5}~\mathrm{Mpc}^{-3}$ for galaxies with
 $M_\star\ge10^{10.5}~M_{\odot}$ and
 $sSFR=0.03$--0.3~$\mathrm{Gyr}^{-1}$, and
 $\sim5.4\times10^{-6}~\mathrm{Mpc}^{-3}$ for LAEs with $EW_0
 (\mathrm{Ly}\alpha)\ge100$~{\AA}).  Namely, the projection
 probability of these objects with similar redshifts (e.g., $\Delta
 z<0.5$) is expected to be extremely low ($\sim0.003$ such chance
 alignments within $1^{\prime\prime}$ in our survey volume).  Another
 idea is that a star-forming dwarf galaxy is going to merge onto these
 MAESTLOs.  In this case, we have to explain why a merging dwarf
 galaxy experiences such active star formation.  Although we cannot
 determine which mechanism is dominant solely using the present data,
 future detailed observations of MAESTLOs such as integral field
 spectroscopy will be useful for this issue.

 How the star formation was quenched in high-redshift massive galaxies
 is now the most important issue for understanding galaxy formation
 and evolution.  Therefore, large volume surveys for such massive
 galaxies with extremely large $EW_0(\mathrm{Ly}\alpha)$ will become
 more important in future.

\acknowledgements

We would like to thank both the Subaru and HST staff for their
invaluable help, and all members of the COSMOS team.  We would also
like to thank the anonymous referee for valuable suggestions and
comments.  We also thank Alex Hagen for kindly providing us with the
information of their LAEs.  This work was financially supported in
part by JSPS (YT: 15340059, 17253001, 19340046, and 23244031; TN:
23654068 and 25707010).


\end{document}